\documentclass[fleqn,nobibnotes]{revtex4}

\usepackage{amsmath}
\usepackage{graphicx}

\def\lsim{\raise0.3ex\hbox{$<$\kern-0.75em\raise-1.1ex\hbox{$\sim$}}}
\def\gsim{\raise0.3ex\hbox{$>$\kern-0.75em\raise-1.1ex\hbox{$\sim$}}}

\begin{document}

\title{A quantum mechanical interpretation for the relativity of the space-time}
\author{E. R. Cazaroto}
\email{ecazaroto@yahoo.com.br}
\affiliation{
Instituto de F\'{\i}sica, Universidade de S\~{a}o Paulo,
C.P. 66318,  05314-970 S\~{a}o Paulo, SP, Brazil\\
}

\begin{abstract}
In special relativity theory the physical quantities are generally expressed as function of the velocity. In the particular case of an extended object, the factor $1/\gamma$ of Lorentz contraction of its length in the direction of motion is written as a function of the velocity of the extended object. The same happens with the Lorentz factor $\gamma$ of time interval dilation of a moving clock. In this paper we show that when $\gamma$ is written as a function of the relativistic energy of the considered object the resulting expressions for the length and for the time interval suggest that the contraction of length as well as the dilation of the time interval are manifestations of the wave behavior of matter.
\end{abstract}

\maketitle

%\tableofcontents

\section{Introduction}

The beginning of last century was marked by a revolution in physics. Two distinct theories, the special relativity theory (SRT) and the quantum mechanics (QM), showed that the human perception of the physical laws is valid only in certain limits. While the first must be used to describe physical phenomena involving velocities comparable with the velocity of light, the second theory is indispensable to describe the microscopic world.

The SRT is based on two postulates. One is that all the inertial frames are equivalent, and the other postulate is that the velocity $c$ of light in the vacuum is the same in all the inertial frames. The first postulate alone does not imply that the time and the space are relative quantities. In the Galilean relativity, for example, all the inertial frames are equivalent, however the space and the time are absolute quantities. The space and the time become relative quantities when one assumes the second postulate of SRT, i.e., that the velocity of light in the vacuum is the same in any inertial frame. This assumption implies that measurements made in different inertial frames are connected with each other through the Lorentz transformations instead of the Galilean ones.

Two interesting consequences of the relativity of space-time are the length contraction of an extended object in the direction of its motion and the dilation of the time interval measured by a moving clock. Suppose, for example, an object with length $l_0$ at rest. According to SRT, in an inertial frame in which this object is moving with uniform velocity $v$ in the direction of its length, an observer will see this same object as having a length $l$ given by:
\begin{equation}
l = \gamma ^{-1} \, l_0 \, ,
\label{length1}
\end{equation}
where:
\begin{equation}
\gamma = \frac{1}{\sqrt{1-v^2/c^2}} \, .
\label{gamma1}
\end{equation}
Therefore, the greater is the velocity $v$ of the object the smaller is its length $l$. Suppose now that observers from an inertial frame are seeing a clock moving with uniform velocity $v$. According to SRT, the time interval $\tau _0$ registered by the moving clock is not equal the time interval $\tau$ registered by the clocks of the observers, but these two quantities are related by:
\begin{equation}
\tau = \gamma \, \tau _0 \, ,
\label{time1}
\end{equation}
with $\gamma$ given by Eq. (\ref{gamma1}). Therefore, the time goes slower for a moving object, i.e., $\tau _0 < \tau$.

In contrast to SRT, the QM was formulated to describe the microscopic world. The postulate by de Broglie establishes that there is a wave associated with any material object, for example there is a wave associated with an electron. According to de Broglie postulate, a material object with well defined moment $p$ is described by a monochromatic plane wave whose wavelength $\lambda$ is given by:
\begin{equation}
\lambda = \frac{h}{p} \, ,
\label{de_broglie1}
\end{equation}
where $h$ is the Plank constant. The frequency $\nu$ of this wave is given by:
\begin{equation}
\nu = \frac{E}{h} \, ,
\label{de_broglie2}
\end{equation}
where $E$ is the energy of the object.

Although there exists the relativistic formulation of QM, in principle QM and SRT are two completely distinct theories. For example, if someone asks to a theoretical physicist if the length contraction and the time interval dilation predicted by SRT have something to do with the wave behavior of matter, the immediate answer certainly will be ``{\it no, nothing to do}''. However, if the theoretical physicist pays a little more attention to this question he will see that there is a qualitative similarity between the de Broglie formula (\ref{de_broglie1}) and the formula of length (\ref{length1}). Eq. (\ref{length1}) tells us that the greater is the velocity $v$ of an object the smaller is its length $l$, whereas Eq. (\ref{de_broglie1}) tells us that the greater is the moment $p$ of the object the smaller is its wavelength $\lambda$. It turns out that the moment is an increasing function of the velocity, so that saying that the velocity increased is equivalent to saying that the moment increased. Consequently, when the velocity of an object increases not only its length decreases but its wavelength also decreases. In other words we can say that the dynamics of the length $l$ of an extended object is {\it qualitatively} similar to the dynamics of the wavelength $\lambda$ associated with the same object, both ($l$ and $\lambda$) being a decreasing function of the object velocity. After this observation, the next step is to investigate {\it quantitatively} the relation between the two dynamics and verify if it is possible to interpret the contraction of length in SRT as being a manifestation of the wave behavior of matter. This is the aim of the present work, and this analysis will be done in Section \ref{length_qm}. In Section \ref{time_qm} we will make a similar analysis for the dilation of the time interval measured by a moving clock. In this analysis we will discuss in details the connection between the time interval $\tau$ registered by a clock and the frequency $\nu$ of the wave associated with the clock. Finally, in Section \ref{conc} we make a summary and the final conclusions.

\section{The spatial length and the wavelength of an extended object}
\label{length_qm}

Consider the problem of an extended object moving with uniform velocity $v$. As discussed in the introduction, the length $l$ of this object in the direction of its motion is given by Eq. (\ref{length1}), with $\gamma$ a function of the object velocity $v$ given by Eq. (\ref{gamma1}). We can, however, rewrite $\gamma$ as a function of the energy $E$ of the object and study the dependence of $l$ on $E$. Let us do it. 

The energy of a relativistic object is given by:
\begin{equation}
E = \frac{mc^2}{\sqrt{1-v^2/c^2}} = \gamma \, E_0 \, ,
\label{energy1}
\end{equation}
where we defined the rest energy of the object as $E_0 = mc^2$. From (\ref{energy1}) $\gamma$ can be written as:
\begin{equation}
\gamma = \frac{E}{E_0} \, ,
\label{gamma2}
\end{equation}
what shows that $\gamma$ is an increasing function of the object energy $E$. Substituting (\ref{gamma2}) in (\ref{length1}) we obtain:
\begin{equation}
l = \frac{l_0 E_0}{E} \, .
\label{length2}
\end{equation}
Therefore, in SRT the length $l$ of an extended object in the direction of motion is inversely proportional to the energy $E$ of the object. This behavior is different of that one of the wavelength $\lambda$ of the object, which according to Eq. (\ref{de_broglie1}) is inversely proportional to the moment $p$ of the object. Let us analyse this difference with more details. In SRT the energy $E$ of an object is related with the moment $p$ through:
\begin{equation}
E = \sqrt{(mc)^2 + p^2} \,\, c \, .
\label{energy2}
\end{equation}
Substituting (\ref{energy2}) in (\ref{length2}) we obtain:
\begin{equation}
l = \frac{l_0 E_0}{\sqrt{(mc)^2 + p^2} \,\, c} \, .
\label{length4}
\end{equation}
For very high moment ($p \gg m\,c \,$) Eq. (\ref{energy2}) implies that $E \approx p \, c$. Consequently, in the limit of very high moment the length $l$ of an extended object, Eq. (\ref{length4}), has the same behavior as the wavelength $\lambda$ of the object, both ($l$ and $\lambda$) being inversely proportional to the moment $p$. However, in the general case of not very high moment, $l$ and $\lambda$ have different behaviors. In particular, in the limit $p \rightarrow 0$ we obtain $E = mc^2$. In this limit the wavelength $\lambda \rightarrow \infty$ whereas the length $l$ acquires the finite value $l = l_0$.

In summary, although the behavior of $l$ has a qualitative similarity with the behavior of $\lambda$, quantitatively the behaviors are different. Therefore, in this first analysis the conclusion is that the length contraction of an extended object in SRT has nothing to do with the wave behavior of matter. However, there is a striking observation to be done. Eq. (\ref{length2}) is telling us that, in SRT, the length $l$ of an object is inversely proportional to the energy $E$ of the object, or conversely that the energy $E$ is inversely proportional to the length $l$:
\begin{equation}
E = \frac{l_0 E_0}{l} \, .
\label{energy3}
\end{equation}
Remember that the energy $E ^{\prime}$ of a photon (or in general of any massless particle) is inversely proportional to the wavelength $\lambda ^{\prime}$ of the photon:
\begin{equation}
E^{\prime} = \frac{h \, c}{\lambda ^{\prime}} \, .
\label{ener_photon}
\end{equation}
Here $h$ is the Plank constant and $c$ is the velocity of light in the vacuum. Therefore, the dependence of the photon energy $E ^{\prime}$ on the photon wavelength $\lambda ^{\prime}$ is identical to the dependence of the energy $E$ of an extended object on the length $l$ of the object. Conversely, we can say that while we have $\, \lambda ^{\prime} \propto 1/E ^{\prime} \,$ for the photon we have $\, l \propto 1/E \,$ for the extended object. 

On one hand, this similarity could be considered a mere coincidence, since the photon is a massless particle whereas the extended object has non-vanishing rest mass. On the other hand, one can think that this similarity is not a mere coincidence, and that there exists some physical scenario in which the behavior $\, \lambda ^{\prime} \propto 1/E ^{\prime} \,$ of the wavelength of massless particles is valid also for massive particles. In this physical scenario, however, the formula by de Broglie, Eq. (\ref{de_broglie1}), should remain valid, since that formula is the basis of QM. The only way of having $\, \lambda \propto 1/E \,$ for massive particles and at the same time having the de Broglie formula (\ref{de_broglie1}) valid is a hypothetical physical scenario in which we can associate to the massive particles the `` {\it effective moment} '':
\begin{equation}
p_{eff} \, = \, \frac{mc}{\sqrt{1-v^2/c^2}} \, ,
\label{p_eff}
\end{equation}
instead of the usual moment:
\begin{equation}
p = \frac{mv}{\sqrt{1-v^2/c^2}} \, .
\label{p_rel}
\end{equation}
With the definition (\ref{p_eff}) for the effective moment, the relativistic energy of a massive particle can be written as:
\begin{equation}
E \, = \, \frac{mc^2}{\sqrt{1-v^2/c^2}} \, = \, p_{eff} \, c \, .
\label{e_eff}
\end{equation}
Consequently, the expression (\ref{length2}) for the length $l$ of an extended object can be rewritten as a function of $p_{eff}$:
\begin{equation}
l = \frac{l_0 E_0}{p_{eff} \, c} \, .
\label{length3}
\end{equation}
Finally, note that by using $\, p_{eff} \,$ in the de Broglie formula (\ref{de_broglie1}) we obtain the following expression for the `` {\it effective wavelength} '' of the extended object:
\begin{equation}
\lambda _{eff} = \frac{h}{p_{eff}} = \frac{h c}{E} \, .
\label{lambda_eff}
\end{equation}

As we can see, if instead of using the usual moment (\ref{p_rel}) in the de Broglie formula (\ref{de_broglie1}) we use the effective moment (\ref{p_eff}), we obtain the effective wavelength (\ref{lambda_eff}) for the massive object, which is inversely proportional to the energy $E$ of the object. In this hypothetical physical scenario we have for a massive object $\, \lambda _{eff} \propto 1/p_{eff} \,$ and $\, l \propto 1/p_{eff} \,$. Consequently, we can interpret the contraction of the object length as being a manifestation of the wave behavior of matter. But how to interpret the ``effective moment''?

The effective moment (\ref{p_eff}) can be rewritten in the following form:
\begin{equation}
p_{eff} = \sqrt{(mc)^2 + p^2} \, .
\label{p_eff2}
\end{equation}
This relation suggests that the term $\, mc \,$ is a component of moment, and that it is perpendicular to the usual relativistic moment $p$. If this is really the case, certainly the component of moment $\, mc \,$ is associated with a microscopic motion, otherwise we would see this ``extra'' motion in our macroscopic world. In fact, there are at least two works proposing that $\, mc \,$ is a term of moment and that it is associated with a microscopic motion. One of these works was written by the present author \cite{the_mass}. In Ref. \cite{the_mass} we proposed that all the massive fundamental particles are doing a microscopic motion with linear moment $\, mc \,$. We analysed a physical scenario in which this motion is a microscopic orbital circular motion (MOCM) and verified that the characteristic length scale of the MOCM is of the order of the Compton wavelength of the particle. Another work proposing to interpret the term $\, mc \,$ as a term of moment which is associated with a microscopic motion was published in \cite{alonso}. In Ref. \cite{alonso} the authors proposed that the microscopic motion (whose moment is $\, mc \,$) takes place in a compacted extra dimension, and showed that this hypothesis offers an explanation for the fact that there are only $3$ generations of fermions in the standard model of particle physics.

Even if both models, the microscopic orbital circular motion analysed in \cite{the_mass} as well as the motion in a compacted extra dimension analysed in \cite{alonso}, are not correct, we should note that the relativistic expression for the energy:
\begin{equation}
E = \sqrt{(mc)^2 + p^2} \, c \, 
\end{equation}
suggests that $\, mc \,$ is a term of moment, and the present paper is giving additional support to this idea by showing that if $\, mc \,$ is really a term of moment we can interpret the contraction of length in SRT as being a manifestation of the wave behavior of matter.

\section{The clock frequency and the wave frequency}
\label{time_qm}

\begin{figure}[htbp]
\includegraphics[scale=0.20]{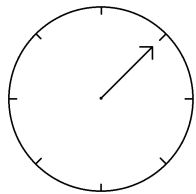}
%\hspace{2.cm}
%\includegraphics[scale=0.50]{fig1b.eps}
%\vspace{0.4cm}
\caption{Hand clock.}
\label{fig1}
\end{figure}

Let us now extend the above analysis for the dynamics of time. We want to verify if it is possible to interpret the dilation of a time interval in SRT as being a consequence of the wave behavior of matter. For this we will compare the Plank relation (\ref{de_broglie2}) with the relativistic expression (\ref{time1}). However, the Plank relation refers to the {\it frequency} $\, \nu \,$ of the wave, so that in order to compare Eq. (\ref{de_broglie2}) with Eq. (\ref{time1}) we must rewrite (\ref{time1}) substituting the time interval $\tau$ registered in a clock by the frequency $f$ of some cyclic physical phenomenon. In this case, the best cyclic phenomenon to be considered is the revolution of the pointer of a hand clock (see Fig. \ref{fig1}). 

According to Eq. (\ref{time1}), the time interval $\tau _0$ registered by a moving clock is smaller than the time interval $\tau$ registered by the clocks of an observer (the observer needs to have at least two clocks, placed in different points along the trajectory of the moving clock, in order to take this conclusion - see e.g. Ref. \cite{relativ}). This means that the frequency `` $f$ '' of revolution of the observer's clock pointer is greater than the frequency `` $f_0$ '' of revolution of the moving clock pointer, i.e., $f>f_0$. To visualize this the reader can think on a situation in which the velocity of the moving clock has a particular value so that Eq. (\ref{time1}) returns $\tau = n \, \tau _0$, with $n$ an integer greater than unity. In this situation, while the pointer of the moving clock completes one revolution, the pointer of the observer's clock complete $n$ revolutions. Therefore, the frequency $f$ of revolution of the observer's clock pointer is always greater than the frequency $f_0$ of revolution of the moving clock pointer. We can then rewrite Eq. (\ref{time1}) making the following substitutions: $\, \tau \rightarrow f \,$ and $\, \tau _0 \rightarrow f_0 \,$. In doing so we obtain:
\begin{equation}
f = \gamma \, f _0 \, .
\label{freq1}
\end{equation}
Using Eq. (\ref{gamma2}) to express $\gamma$ in terms of the relativistic energy we obtain:
\begin{equation}
f = \frac{E}{E_0} \, f _0 \, .
\label{freq2}
\end{equation}
Before discussing the meaning of this equation let us make the following observation. The Plank relation (\ref{de_broglie2}) tells us that the greater is the energy $E$ of an object the greater is the frequency $\nu$ of the wave associated with the object. So, in a first moment one could conclude that the frequency $f_0$ of revolution of the moving clock pointer should be proportional to the energy $\, E \,$ of the moving clock, i.e., $\, f_0 \propto E \,$, in order that we can interpret the time interval dilation in SRT as being a manifestation of the wave behavior of matter. However, what we have in Eq. (\ref{freq2}) is the opposite of this: for fixed $f$ the greater is the energy $E$ the smaller is the frequency $f_0$. What is wrong?

The mistake in the above reasoning is that the quantity $\tau _0$ (and consequently $f_0$) is a Lorentz invariant quantity. In what follows we will explain this better.

\begin{figure}[htbp]
\includegraphics[scale=0.40]{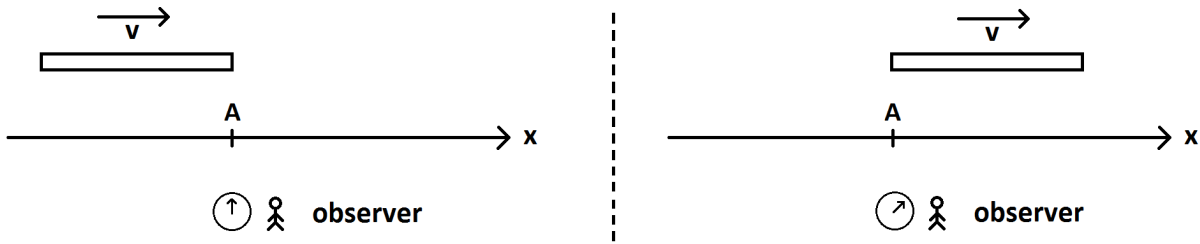}
%\hspace{2.cm}
%\includegraphics[scale=0.50]{fig1b.eps}
%\vspace{0.4cm}
\caption{An observer measuring the length of an extended object that is moving with velocity $v$ along the $x$ axis.}
\label{fig2}
\end{figure}

There is a crucial difference between measuring the length of a moving extended object and comparing the time interval of the observer's clock with the one of a moving clock. The extended object has two extremes, so that the observer needs one only point `` $A$ '' of reference in his inertial frame in order to measure the length of the object (see Fig. \ref{fig2}). The observer knows the velocity $v$ of the object. When the right extreme of the object reaches the point $A$ the observer looks at his clock (Fig. \ref{fig2} - left). When the left extreme of the object reaches the point $A$ the observer looks again at his clock (Fig. \ref{fig2} - right). According to SRT, the observer will conclude that the length $l$ of the object is:
\begin{equation}
l = v \, \Delta t \, ,
\end{equation}
where $\Delta t$ is the time interval (measured by the observer in his clock) elapsed between the situation in the Fig. \ref{fig2} - left and the situation in the Fig. \ref{fig2} - right. Note that there is a spatial asymmetry in this measurement procedure. The observer needs one only spatial point `` $A$ '' in his referential to make the measurement, whereas the extended object ``contributes'' with two spatial points in this process, namely the right and the left extremes of the object. 

\begin{figure}[htbp]
\includegraphics[scale=0.40]{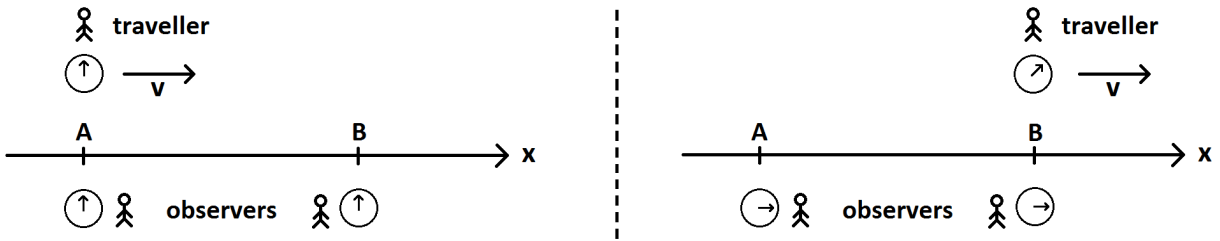}
%\hspace{2.cm}
%\includegraphics[scale=0.50]{fig1b.eps}
%\vspace{0.4cm}
\caption{Two observers comparing the time in their (synchronized) clocks with the time in the clock of a traveller moving with velocity $v$ along the $x$ axis.}
\label{fig3}
\end{figure}

The comparison of time intervals is different in the following sense. The moving clock is not an extended object, so that it is considered punctual. Therefore, according to SRT it is necessary to have two observers, placed in different points $A$ and $B$, with synchronized clocks, to compare the time registered in their clocks with the time registered in the moving clock (see Fig. \ref{fig3}). When the moving clock reaches the point $A$ (Fig. \ref{fig3} - left) the observer placed in this point compares the time in his clock with the time in the moving clock. When the moving clock reaches the point $B$ (Fig. \ref{fig3} - right) the observer placed in this point compares the time in his clock with the time in the moving clock. After informing each other about their observations, the observers will conclude that the time goes slower for the moving clock. 

Now consider a traveller moving together with the clock, as shown in Fig. \ref{fig3}. The traveller also compares the time in the moving clock with the time in the clocks of the observers when passing through the points $A$ (Fig. \ref{fig3} - left) and $B$ (Fig. \ref{fig3} - right). The time interval $\tau _0$ registered by the traveller's clock while goes from $A$ to $B$ is smaller than the time interval $\tau$ registered by the observer's clock. Consequently, the frequency $f$ of revolution of the observer's clock pointer is greater than the frequency $f_0$ of revolution of the traveller's clock pointer. Therefore, the traveller will conclude that the time goes faster for the observers. This happens because the traveller is using one only spatial point of his referential (namely the position of his clock) to make the comparisons, whereas the observers ``contribute'' with two spatial points in this process. Note that the situation of the traveller from Fig. \ref{fig3} is similar to the situation of the observer from Fig. \ref{fig2} in the sense that this last one also uses one only point of his referential (namely the position of his clock) to make the measurement of the time interval.

Now, returning to the analysis of Eq. (\ref{freq2}) we understand that it is the frequency $f$ (associated with the clocks of the observers) that is proportional to the energy $E$ and not the proper frequency $f_0$ of the moving clock. This last quantity is Lorentz invariant, i.e., it is invariant under change of referential. Furthermore, note that the ratio $E/E_0$ corresponds to the factor $\gamma$, which in turn depends on the square of the velocity. It means that the factor $\gamma$ calculated by the observers considering the velocity $+v$ of the moving clock is the same factor $\gamma$ obtained by the traveller by considering the velocity $-v$ of the clocks of the observers. Therefore, the traveller can interpret the energy $E$ appearing in Eq. (\ref{freq2}) as being the total energy of the two clocks of the observers. Consequently, the traveller will conclude that the frequency $f$ of revolution of the observer's clock pointer is proportional to the energy $E$ of the observer's clock, i.e., $f \propto E$. According to the Plank relation (\ref{de_broglie2}), this is exactly the dependence of the frequency $\nu$ of the wave associated with the clocks on the energy $E$ of the clocks, i.e., $\nu \propto E$. Therefore, from the point of view of the traveller the relativity of a time interval (an effect predicted by SRT) can be interpreted as being a manifestation of the wave behavior of matter.

\section{Summary}
\label{conc}

We showed that it is possible to interpret the contraction of length as well as the dilation of a time interval (two phenomena predicted by SRT) as being manifestations of the wave behavior of matter. For the contraction of length this interpretation depends of considering a hypothetical physical scenario in which the effective moment (\ref{p_eff2}) is inserted in the de Broglie formula to determine the wavelength of the wave associated with a material object. The ``effective moment'' defined in Eq. (\ref{p_eff2}) indicates that the term $\, mc \,$ is a term of moment, which is associated with some microscopic motion. For the dilation of a time interval we do not need to consider any hypothesis. The interpretation is immediate. To take this conclusion we considered the example in which the frequency $f$ of revolution of the observer's clock pointer is compared with the proper frequency $f_0$ of revolution of the traveller's clock pointer. We verified that $f$ is proportional to the energy $E$ of the observer's clock. Consequently, the traveller can interpret the relativity of the time interval as being a manifestation of the wave behavior of matter.

\end{document}